\documentclass[
  twocolumn,
  prb,
  amsmath,
  amssymb,
  superscriptaddress,
  floatfix
]{revtex4}

\usepackage{graphicx}

\renewcommand{\cite}[1]{{[}\onlinecite{#1}{]}}

\newcommand{\s}{\sum\limits}

\newcommand{\il}{\int\limits}
\newcommand{\be}{\begin{equation}}
\newcommand{\e}{\end{equation}}
\newcommand{\bbul}{\begin{multline}}
\newcommand{\emul}{\end{multline}}
\newcommand{\beml}{\begin{subequations}}
\newcommand{\eml}{\end{subequations}}
\newcommand{\beq}{\begin{eqnarray}}
\newcommand{\eq}{\end{eqnarray}}
\newcommand{\ba}{\begin{array}}
\newcommand{\ea}{\end{array}}
\newcommand{\bpm}{\begin{pmatrix}}
\newcommand{\epm}{\end{pmatrix}}
\newcommand{\bc}{\begin{cases}}
\newcommand{\ec}{\end{cases}}
\newcommand{\lt}{\left}
\newcommand{\rt}{\right}
\newcommand{\n}{\nonumber}
\newcommand{\la}{\langle}
\newcommand{\ra}{\rangle}
\newcommand{\ep}{\varepsilon}

\newcommand{\bb}{\boldsymbol}

\DeclareMathOperator{\tr}{Tr}

\begin{document}

\title{Sensitivity of anomalous Hall effect to disorder correlations}

\author{I.\,A.~Ado}
\affiliation{Radboud University, Institute for Molecules and Materials, NL-6525 AJ Nijmegen, The Netherlands}
\author{I.\,A.~Dmitriev}
\affiliation{University of Regensburg, Department of Physics, 93040 Regensburg, Germany}
\affiliation{Max Planck Institute for Solid State Research, Heisenbergstr.\,1, 70569 Stuttgart, Germany}
\affiliation{A.\,F.~Ioffe Physico-Technical Institute, 194021 St.\,Petersburg, Russia}
\author{P.\,M.~Ostrovsky}
\affiliation{Max Planck Institute for Solid State Research, Heisenbergstr.\,1, 70569 Stuttgart, Germany}
\affiliation{L.\,D.~Landau Institute for Theoretical Physics RAS, 119334 Moscow, Russia}
\author{M.~Titov}
\affiliation{Radboud University, Institute for Molecules and Materials, NL-6525 AJ Nijmegen, The Netherlands}
\affiliation{ITMO University, Saint Petersburg 197101, Russia}

\begin{abstract}
Both longitudinal and anomalous Hall conductivity are computed in the model of two-dimensional Dirac fermions with a mass in the presence of arbitrary correlated weak disorder. The anomalous Hall conductivity is shown to be highly sensitive to the correlation properties of the random potential, such as the correlation length, while it remains independent of the integral disorder strength. This property extends beyond the Dirac model making the anomalous Hall effect an interesting tool to probe disorder correlations.
\end{abstract}
%\pacs{72.10.-d, 72.25.-b, 72.10.Bg}

\maketitle

\section{Introduction}

\noindent
The anomalous Hall effect (AHE) is one of the most direct manifestations of spin-orbit interaction in magnetic conductors. The effect has been discovered as early as in 1881 by Edwin Hall who observed a transverse voltage in ferromagnetic iron as a reaction to electric current applied \cite{Hall81}. In that respect the AHE is completely analogous to the usual Hall effect but can be observed in much weaker magnetic fields that are only needed to magnetize the conductor \cite{Nagaosa10}. A closely related phenomenon of the AHE in antiferromagnetic and paramagnetic systems has, however, received a widespread attention only recently \cite{Kubler14,Nakatsuji15,Maryenko17}.

The interest to spin-orbit induced phenomena has increased dramatically following the discovery of topological insulators and Weyl semimetals \cite{Hasan2010rev, Qi2011rev, Xu15, Xiong15, Kim15}. Moreover, the on-going development in the fields of spintronics \cite{Pesin2012,Chappert07, Wunderlich10, Jungwirth12}, cold atoms \cite{Lin11, Cheuk12, Jotzu14}, chiral superconductivity \cite{Xia2006, Xia2008, Kapitulnik2009, Schemm2014}, and magnetization dynamics \cite{Garello2013, Nagaosa2013, Melnik2014, Fan2014, Liu15} call for deeper understanding of the microscopic mechanisms of spin-orbit assisted transport \cite{Manchon15, Sinova15, Hoffmann13}. The detailed interpretation of the AHE measurements may provide valuable information regarding exchange and spin-orbit coupling that is of key importance for applications.

Despite long and rich history of the field \cite{Nagaosa10}, it has been recently found by the authors \cite{Ado15,Ado16} that previous treatments of AHE were fundamentally incomplete. Indeed, in many models and materials the AHE conductivity is sub-leading in a large metal parameter $\ep_F\tau$ compared to the longitudinal conductivity (here $\ep_F$ is the Fermi energy). This is reflected in the fact that the anomalous Hall conductivity does not depend on the electron scattering time $\tau$ and is of the order of intrinsic contribution which is manifestly disorder-independent. It appears, nevertheless, that the presence of impurities essentially modify the AHE including its sign \cite{Ado16}, even though the disorder strength is canceling out from the result. Very recently, the same crossed diagrams have been shown to play a key role for the AHE on the  surface of topological Kondo insulators \cite{Koenig16}, for the Kerr effect in chiral p-wave superconductors \cite{Koenig17}, and for the spin Hall effect in the presence of strong impurities \cite{Milletari2016}.

Indeed, it has been demonstrated in Refs.~\cite{Ado15,Ado16} that the conventional ``non-crossing approximation'' (NCA) employed in diagrammatic calculation of longitudinal conductivity $\sigma_{xx}$ is not applicable to AHE.  In addition to ladder diagrams with non-crossing impurity lines that describe electron diffusion, the AHE conductivity $\sigma_{xy}$ requires additional terms that are represented by diagrams with two intersecting impurity lines (the so-called X and $\Psi$ diagrams \cite{Ado15,Ado16}). From physics point of view such terms represent an essential part of the full cross section describing skew scattering on pairs of closely positioned impurities. The important role of such rare impurity configurations in the theory of AHE calls for a detailed investigation of the effects of disorder correlations that is the main subject of the present publication.  

For two basic microscopic models of AHE \cite{Nagaosa10} the anomalous Hall conductivity $\sigma_{xy}$ has been shown to depend dramatically on the inclusion of X and $\Psi$ diagrams \cite{Ado15,Ado16}. More specifically, in 2D Rashba ferromagnet the AHE conductivity does not vanish in the metallic limit solely due to these contributions  \cite{Ado16} in sharp contrast to the well-known vanishing NCA result \cite{Inoue06,Nunner07}. In the case of massive Dirac fermions, which represent the simplest model featuring the AHE, the X and $\Psi$ diagrams almost cancel out the NCA contribution \cite{Ado15}. In the model with weak white noise disorder the anomalous Hall conductivity $\sigma_{xy}$ decays as $\ep_F^{-3}$ in the metal regime instead of $\ep_F^{-1}$ given by the NCA \cite{Sinitsyn2006,Sinitsyn2007}.

Given that the AHE is so sensitive to the scattering on rare impurity configurations it is interesting to establish whether and how these results modify for more general disorder models with finite correlation length. Here we extend the analytic approach of Refs.~\cite{Ado15, Sinitsyn2006} to massive Dirac fermions subject to weak Gaussian disorder with arbitrary pair correlator of the random potential.

As might have been expected our results reveal strong sensitivity of the AHE conductivity to the correlation properties of disorder. In particular, it turns out that strong mutual cancellation of intrinsic and extrinsic contributions for the model of Dirac fermions \cite{Ado15} is the specific feature of uncorrelated disorder. In the opposite limit of smooth disorder, that correspond to small-angle scattering, the intrinsic and extrinsic contributions to $\sigma_{xy}$ have the same sign. In this case the total Hall conductivity at large energies $\sigma_{xy}\propto\ep_F^{-1}$ is given by the intrinsic conductivity multiplied by a factor of three. Depending on the functional form of the pair correlator, the AHE conductivity may feature non-monotonic dependence on the correlation length and possess maximal values in crossover region between the above limits of white-noise and smooth disorder.

The paper is organized as follows. In Sec.~\ref{secModel}, we introduce the model and calculate the disorder-averaged Green's functions. In Sec.~\ref{secGeneral} we obtain general expressions for the longitudinal and anomalous Hall conductivities in terms of the angular moments of the disorder correlation function.  In Sec.~\ref{secLimits} these general results are applied to generic limiting cases of white-noise and smooth disorder. Section \ref{secCrossover} illustrates the crossover between the two limits for two specific models of disorder. Section \ref{secSummary} contains summary and conclusions. Certain technical details of calculations and complementary information are presented in Appendices \ref{app_derivation} and \ref{app_separate}.

\section{Model, disorder-averaged Green's functions}
\label{secModel}

\subsection{Model}

The model of massive Dirac fermions in two dimensions has been proposed by Haldane \cite{Haldane1988} as the simplest toy model to illustrate the quantum anomalous Hall effect. The latter arises when the chemical potential is placed in the band gap. In this paper we focus on the metal regime, i.\,e. on the case of chemical potential situated within the band. In particular we compute both longitudinal and Hall conductivity for the model of massive Dirac fermions in two dimensions that is described by the Hamiltonian
\be
\label{H}
H  = H_0+V(\bb{r}),\qquad H_0=v \bb{\sigma} \bb{p} + m \sigma_z,
\e
where $V(\bb{r})$ denotes a weak correlated Gaussian random potential, $\bb{\sigma}=(\sigma_x,\sigma_y)$ stands for the vector of Pauli matrices, $\bb{p}=(p_x, p_y)$ is the momentum operator, $v$ is the characteristic velocity and $m$ is the bare ``mass'' of relativistic fermions. The random potential is characterized by a correlator
\be
\label{V}
\la V(\bb{r}) V(\bb{r}') \ra =  2\pi \widetilde\alpha(\bb{r} - \bb{r}'),
\e
where the angular brackets denote the averaging over the disorder realizations. In diagrammatic language such a correlator is visualized by a disorder line. Each line
corresponds to the propagator $2\pi \alpha(\bb{q})=2\pi\int d^2\bb{r}\, \widetilde{\alpha}(\bb{r})e^{-i\bb{q}\bb{r}}$ which depends on the transferred momentum $\bb{q}$. Throughout the paper we use  $\hbar=1$ and $v=1$, hence all momenta are measured in the units of energy.

The model of Eq.~(\ref{H}) is characterized by the broken time-reversal invariance hence it gives rise to a finite Hall response. In condensed matter context the Hamiltonian of the type (\ref{H}) can be used as an effective model \cite{Fu2009,Garate2010,Yokoyama2010,Tserkovnyak2012} to describe the surface of the 3D topological insulator \cite{Chen2010,Melnik2014} (see, however, Ref.~\onlinecite{Menshov2013} for the detailed discussion). One may also view the model of Eq.~(\ref{H}) as the single-valley projection of the full Hamiltonian describing graphene/hBN heterostructure \cite{Woods2014}. In the latter case, however, the Pauli matrices $\sigma_\alpha$ act in the isospin space while the true time-reversal invariance of the Hamiltonian is preserved. This means that the AHE for the model of Eq.~(\ref{H}) corresponds to the valley Hall effect in graphene/hBN bilayer.

The quantum anomalous Hall effect, studied by Haldane \cite{Haldane1988} in the model of Eq.~(\ref{H}), is manifestly independent of the disorder potential and is taking place for the chemical potential placed within the band gap. To study transport properties outside the gap one needs to take into account the scattering on impurities both in the longitudinal and in the anomalous Hall conductivity \cite{Ludwig1994}. The missing leading-order terms in the theory of AHE have been discovered by the authors only recently \cite{Ado15}.

In what follows we compute the components of the conductivity tensor to the leading order in $\alpha$. In order to do so, it is sufficient to know the disorder scattering
probability only for the states belonging to the Fermi surface. The spectrum of $H_0$ consists of two branches $\ep_\pm(p)=\pm\sqrt{p^2 + m^2}$ separated by a gap of the size $2|m|$. Without loss of generality we assume that the Fermi energy $\ep$ belongs to the upper band $\ep>m>0$ so that $p_0 = \sqrt{\ep^2 - m^2}$ is the corresponding Fermi momentum. The transferred momentum $q = 2 p_0 \sin(\phi/2)$ is, then, uniquely expressed by the scattering angle $\phi$. With these definitions we express the angle-dependent scattering probability as
\be
\label{alphaphi}
\alpha(\phi) \equiv \alpha[q\to 2 p_0 \sin(\phi/2)]  = \alpha_0 + 2 \sum_{n = 1}^\infty \alpha_n \cos(n \phi),
\e
where the parameters $\alpha_n$ stand for the angular harmonics of the scattering probability. The notation $\alpha(\phi)$ is used interchangeably with $\alpha(q)$ below. The particular limit of white-noise disorder, $\widetilde{\alpha}(\bb{r})=\alpha \delta(\bb{r})$, which has been investigated in Ref.~\onlinecite{Ado15}, corresponds to $\alpha_n=\alpha\delta_{n,0}$, where $\delta_{n,m}$ is the Kronecker symbol.

\subsection{Self energy and average Green's functions}

The main building block of diagrammatic calculations below is the Green's function averaged over disorder configurations in the Born approximation.
The latter is defined by the corresponding self-energy with a logarithmically diverging real part, which is absorbed in the renormalization of energy and mass, and with a finite imaginary part. The latter is set by the difference between the retarded and advanced self-energy
\be
\label{sigma}
\Sigma^R_{\bb{p}} - \Sigma^A_{\bb{p}} = \int \frac{d^2 \bb{p}'}{2\pi}\, \alpha(\bb{p} - \bb{p}') \lt[ G^R(\bb{p}') - G^A(\bb{p}') \rt],
\e
where the Green's functions can be taken in the clean limit. The bare Green's functions [which correspond to the Hamiltonian $H_0$ in the Eq.~(\ref{H})] yield
\be
G^R_0(\bb{p}) - G^A_0(\bb{p})  = -2 \pi i [\ep + m \sigma_z + \bb{\sigma} \bb{p}] \delta(p^2 - p_0^2),
\e
where the presence of delta-function bounds the particle momentum $\bb{p}$ to the Fermi surface.

We only need to know the self-energy for momenta at the Fermi surface, $p = p_0$, consequently we find the result
\begin{align}
&\Sigma^R_{\bb{p}}- \Sigma^A_{\bb{p}}  = -i \pi \int \frac{d\phi'}{2\pi}\, \alpha(\phi - \phi')
\lt(\ep + m \sigma_z + p_0 \sigma_x \cos\phi'\rt.\n\\
&\qquad\lt.+ p_0 \sigma_y \sin\phi' \rt)= -i \pi \lt(\alpha_0 (\ep + m \sigma_z) + \alpha_1 \bb{\sigma} \bb{p} \rt),
\label{se}
\end{align}
which depends only on the first two harmonics of the disorder correlator.

Using the self energy of Eq.~(\ref{se}) we obtain the averaged Green's function in the form
\be
G^{R,A}(\bb{p})
= \frac{\ep \pm i \gamma + (m \mp i \mu) \sigma_z + (1 \mp i \zeta) \bb{\sigma} \bb{p}}{\ep^2 - m^2 - p^2 \pm i \Gamma},
\label{full}
\e
where we introduce the parameters
\begin{align}
  \n
 &\Gamma
  = 2(\ep \gamma + m \mu + \zeta p^2)
  = \pi \lt( \ep^2 (\alpha_0 + \alpha_1) + m^2 (\alpha_0 - \alpha_1) \rt),
  \\
  &\gamma
  = \pi \alpha_0 \ep/2,
 \qquad
 \mu
  = \pi \alpha_0 m/2,
 \qquad
 \zeta
  = \pi \alpha_1/2,
\end{align}
with the expression for $\Gamma$ taken at $p = p_0$. This is justified since the imaginary part of the Green's function denominator is relevant only at the mass shell.

The singular part of the average Green function comes from the Fermi surface and can be obtained from Eq.~(\ref{full}) via projection on the corresponding spectral branch. 
For $\ep>m$ it is given by
\be
\label{proj}
G^{R,A}_+(\bb{p})  = \frac{|\phi \ra \la \phi|}{\ep - \sqrt{m^2 + p^2} + i/2\tau},
 \e
where we introduced the scattering rate $1/\tau=\Gamma|_{p=p_0}/\ep$ and the eigenstate
\be
\label{state}
|\phi \ra  = \frac{1}{\sqrt{2\ep}} \bpm \sqrt{\ep + m} \\ \sqrt{\ep - m}\, e^{i\phi} \epm,
\e
which corresponds to the momentum (taken at the Fermi surface) pointing out in the direction $\phi$.

The scattering rate $1/\tau$ in Eq.~(\ref{proj}) is expressed by the Fermi golden rule as
\be
\label{tau}
1/{\tau} = 2 \pi \ep\, \lt[\alpha(\phi) \Delta(\phi)\rt]_{\phi},
\quad \Delta(\phi - \phi') \equiv \lt| \la \phi | \phi' \ra \rt|^2,
\e
where the square brackets denote the angular averaging
\be
\lt[ u(\phi) \rt]_\phi \equiv \int\frac{d\phi}{2\pi}\, u(\phi),
\e
and we introduce the so-called Dirac factor
\be
\label{delta}
\Delta(\phi)
  = \cos^2 \frac{\phi }{2} + \frac{m^2}{\ep^2} \sin^2 \frac{\phi}{2},
\e
which reflects the structure of the eigen states. The expressions (\ref{sigma}--\ref{delta}) provide the basis for the diagrammatic analysis of the conductivity tensor which we undertake in the next section.

\section{Conductivity tensor for correlated weak Gaussian disorder}
\label{secGeneral}

\subsection{General remarks}

To compute the dc conductivity tensor for the system described by Eqs.~(\ref{H}--\ref{alphaphi}) we employ the Kubo-Streda formula \cite{Streda} in the limit of  zero temperature and for the Fermi energy, $\ep>m>0$, belonging to the conduction band. In this paper we generalize the formalism developed in Ref.~\onlinecite{Ado15} to the case of correlated weak disorder. Unlike Ref.~\onlinecite{Ado15}, where the computation has been performed in the real space representation for the case of uncorrelated (white noise) disorder, we use here the momentum representation, which is more appropriate for dealing with correlated disorder.

The Kubo-Streda formula for the conductivity tensor consists of two terms traditionally denoted as $\hat\sigma^\text{I}$ and $\hat\sigma^\text{II}$. The first term, $\hat\sigma^\text{I}$, describes the contribution of conduction electrons with momenta at the Fermi surface. The second term accounts for the contribution to the non-diagonal components of the conductivity tensor $\sigma_{xy}=-\sigma_{yx}$ that stems from the entire Fermi sea. In particular, the second contribution can be expressed as $\sigma_{xy}^\text{II}= e c\, {\partial n}/{\partial B}|_{B\to 0}$ as the derivative of the total electron concentration $n$ with respect to an external perpendicular magnetic field $B$ taken in the limit $B\to 0$ \cite{Streda}. 

For the Fermi energy inside the spectral gap, $|\ep| < m$, the longitudinal conductivity $\sigma_{xx}$ vanishes, while the entire Hall conductivity is given by  \cite{Haldane1988, Ludwig1994, Ostrovsky2007},
\be
\label{II}
\sigma_{xy}=\sigma_{xy}^\text{II} = -e^2/4\pi,
\e
that remains insensitive to a weak disorder. The result of Eq.~(\ref{II}) is often referred to as the quantum anomalous Hall effect.% One can also show that within the gap the contribution $\sigma_{xy}^\text{II}$ is of topological origin and is directly related to Berry curvature. 

As the Fermi energy is increased above the gap, $\ep>m$, the contribution $\sigma_{xy}^\textrm{II}$ quickly becomes negligible in comparison to the Fermi surface contribution $\sigma_{xy}^\textrm{I}$. In this case the conductivity tensor is given by
\be
\label{I}
\sigma_{ij}^\text{I}=\frac{e^2}{2\pi} \tr \la \sigma_i G^R \sigma_j G^A \ra,
\e
where $G^A$ and $G^R$ stand for the exact Green's functions in the presence of disorder.  The angular brackets denote the averaging over disorder realizations defined by Eq.~(\ref{V}). Let us remind that in our units ($\hbar=1$ and $v=1$) the conductance quantum $e^2/h$ reads $e^2/2\pi$, while the components of the current operator are given by the Pauli matrices $\sigma_{x,y}$. In the following, we compute the conductivity tensor (\ref{I}) as a function of the angular harmonics $\alpha_n$ of the disorder potential correlator given in Eq.~(\ref{alphaphi}) to the leading order in the disorder strength.

\begin{figure}
 \includegraphics[width=\columnwidth]{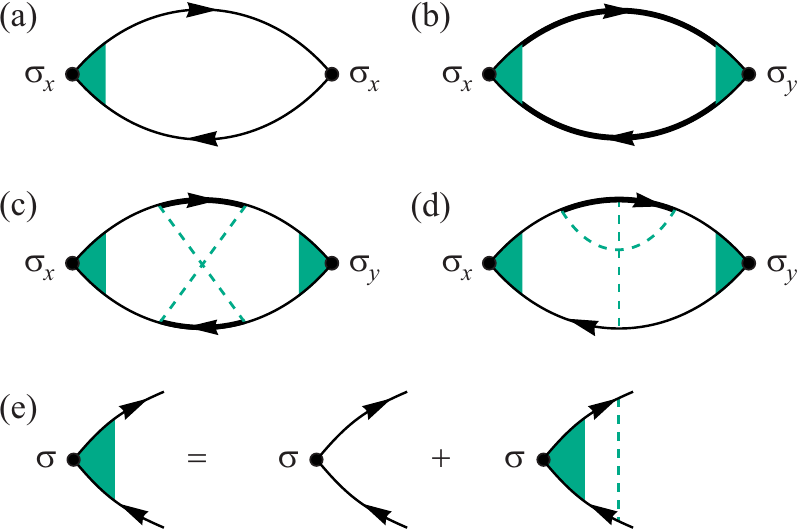}\\
 \caption{Diagrams representing the leading contributions to the longitudinal conductivity $\sigma_{xx}={\cal O}(\alpha_n^{-1})$ [Panel (a)] and the anomalous Hall conductivity $\sigma_{xy}={\cal O}(\alpha_n^{0})$ [Panels (b), (c), and (d)]. The latter is the sum of the non-crossing contribution $\sigma_{xy}^\text{nc}$ [Panel (b)] and crossing contributions represented by X [Panel (c)] and $\Psi$ [Panel (d)] diagrams, which include a pair of crossed impurity lines (dashed lines). Thin solid lines correspond to Green's functions (\ref{proj}) projected to the Fermi surface, while thick solid lines correspond to full disorder-averaged Green's functions (\ref{full}). Vertex correction (e) involves the sum of ladder diagrams with projected Green functions $G_+$ only.
\label{diagrams}
}
\end{figure}

We start in Sec.~\ref{subsecxx} with the longitudinal conductivity $\sigma_{xx}=\sigma_{yy}={\cal O}(\alpha_n^{-1})$ that is proportional to the scattering time $\tau$. The averaging procedure in this case is reduced to the computation of the standard ladder diagrams with non-crossing impurity lines as illustrated in Fig.~\ref{diagrams}(a) and (e). The dominant contribution to $\sigma_{xx}$ is determined entirely by the states belonging to the Fermi surface. To perform the calculation in the leading order $\alpha_n^{-1} \propto \tau$ it is sufficient to know only the singular part of the Green's functions of Eq.~(\ref{proj}), which is represented by thin lines in Fig.~\ref{diagrams}.

The same approximation applied to the anomalous Hall conductivity $\sigma_{xy}$ returns a vanishing result. This is intuitively clear since the full projection to one of the bands restores the time-reversal symmetry of the model. To obtain a finite result for $\sigma_{xy}$ one needs to take into account states that lay far from the Fermi surface. As the result, the AHE has a parametric smallness $\sigma_{xy}={\cal O}(\alpha_n^{0})$ as compared to $\sigma_{xx}={\cal O}(\alpha_n^{-1})$. 
A part of $\sigma_{xy}$ comes from non-crossing diagrams in Fig.~\ref{diagrams}(b), where, in comparison to Fig.~\ref{diagrams}(a), any single pair of the projected Green functions (\ref{proj}) has to be replaced by full Green functions (\ref{full}) (thick lines in Fig.~\ref{diagrams}) which include contribution of states far away from the Fermi surface. The corresponding part of $\sigma_{xy}$ is calculated in Sec.~\ref{subsecNC}.

The non-crossing diagrams in Fig.~\ref{diagrams}(b) do not exhaust all contributions to the AHE $\sigma_{xy}$ even in the leading order with respect to the disorder strength. As demonstrated in Ref.~\onlinecite{Ado15}, a complete description requires inclusion of additional diagrams in Fig.~\ref{diagrams}(c) and (d) which involve a single pair of crossed impurity lines. It is well known that crossing of impurity lines leads to a parametric smallness since the momentum conservation law does not enable bounding of all momenta to the Fermi surface. But in the case of anomalous Hall conductivity one of the momenta needs to be away from the Fermi surface even in the non-crossing diagrams, Fig.~\ref{diagrams}(b). As discussed in more details in Refs.~\cite{Ado15, Ado16, Koenig16, Koenig17}, in this situation crossing of impurity lines in Fig.~\ref{diagrams}(c) and (d) does not produce any additional smallness with respect to Fig.~\ref{diagrams}(b). The corresponding contribution to $\sigma_{xy}$ is calculated in Sec.~\ref{subsecXPsi}.

%%%%%%%%%%%%%%
%Fig 1
%%%%%%%%%%%%%%
%%%%%%%%%%%
%%%%%%%%%%%%%%

\subsection{Longitudinal conductivity}
\label{subsecxx}

The longitudinal conductivity $\sigma_{xx}$ in the leading order in $\alpha_n$ is given by the sum of ladder diagrams illustrated in Fig.~\ref{diagrams}(a). The result can be written as
\be
\label{sxx1}
\sigma_{xx}
= \frac{e^2}{2\pi}
\il_0^{2\pi} \frac{d\phi}{2\pi}\il_0^\infty \frac{p\,dp}{2\pi} \tr \lt( G^A_+(\bb{p}) \bar\sigma_x G^R_+(\bb{p}) \sigma_x \rt),
\e
where $\bar\sigma_x$ represents the current operator, dressed by the impurity ladder as shown in Fig.~\ref{diagrams}(e). In what follows we often do not specify the integration limits explicitly since they are always the same as in Eq.~(\ref{sxx1}).

As have been already mentioned, the calculation of the longitudinal conductivity to the leading order in disorder strength can be preformed entirely with projected Green's functions given by Eq.~(\ref{proj}). This is not the case for the AHE as will be explained in the next subsection.

Let us first compute the matrix element $\la \phi | \bar\sigma_x | \phi \ra$, which represents the disorder-dressed current vertex. The projected bare current operator reads
\be
\la \phi | \sigma_x | \phi \ra= \frac{p_0}{\ep}\, \cos \phi,
\e
where the ratio $p_0/\ep$ is the Fermi velocity. This matrix element is transformed as
\begin{align}
\frac{p_0}{\ep}\, \cos \phi \mapsto &\frac{p_0}{\ep} \int\!\!\!\int \frac{p\, dp\, d\phi'}{2\pi}\,
\frac{\la \phi | \phi' \ra \alpha(\phi - \phi') \la \phi' | \phi \ra\cos \phi'}{(\ep - \sqrt{m^2 + p^2})^2 + 1/4\tau^2} \n \\
&= p_0 \tau \int d\phi'\, \alpha(\phi - \phi') \Delta(\phi - \phi') \cos \phi'\n\\
&= \frac{p_0}{\ep}\, \cos \phi\, \frac{\lt[ \alpha(\phi') \Delta(\phi') \cos \phi' \rt]_{\phi'}}{\lt[\alpha(\phi') \Delta(\phi') \rt]_{\phi'}},
\end{align}
after dressing by a single impurity line. Summing up the entire ladder results in the fully dressed projected current,
\begin{align}
\label{projVertex}
\la \phi | \bar\sigma_x | \phi \ra &
= \frac{p_0}{\ep}\, \cos \phi \frac{\lt[\alpha(\phi') \Delta(\phi') \rt]_{\phi'}}{\lt[ \alpha(\phi') \Delta(\phi') ( 1 - \cos \phi' ) \rt]_{\phi'}}\n\\
  &= \frac{p_0 \tau_\text{tr}}{\ep \tau} \cos\phi,
\end{align}
that defines the transport scattering rate
\begin{align}
\frac{1}{\tau_\text{tr}} &= 2 \pi \ep \lt[ \alpha(\phi) \Delta(\phi) (1 - \cos\phi) \rt]_{\phi}\n\\
&= \frac{\pi}{2 \ep}\, \lt( \ep^2 (\alpha_0 - \alpha_2) + m^2 (3 \alpha_0 - 4 \alpha_1 + \alpha_2) \rt),
\label{ttr}
\end{align}
which depends on the first three harmonics of the random potential. We shall remind here that the scattering time $\tau$ is the average time between two scattering events, while the transport time $\tau_\text{tr}$ is the characteristic time of momentum relaxation. In the limit of very smooth disorder, when $\alpha_n$ is a slow function of $n$, the transport time may exceed the scattering time by orders of magnitude since the small angle scattering dominates.

It is instructive to define the following object
\begin{align}
\label{J}
J_x(\phi) &= \int \frac{p\, dp}{2\pi}\, G^A_+(\bb{p}) \bar\sigma_x G^R_+(\bb{p})\n\\
&= \frac{p_0 \tau_\text{tr}}{\ep \tau} \int \frac{p\, dp}{2\pi}\, \frac{| \phi \ra \cos\phi \la \phi |}{(\ep - \sqrt{m^2 + p^2})^2 + 1/4\tau^2}\n\\
&= p_0 \tau_\text{tr} | \phi \ra \cos\phi \la \phi |,
\end{align}
which is obtained by adding one extra pair of Green's functions to the dressed current (\ref{projVertex}) and integration over the absolute value of the momentum. The result of Eq.~(\ref{J}) helps evaluating Eq.~(\ref{sxx1}) as
\begin{align}
\sigma_{xx} &= \frac{e^2}{2\pi}  \tr \lt[ J_x(\phi) \sigma_x \rt]_\phi = \frac{e^2 \tau_\text{tr}}{4 \pi \ep}(\ep^2 - m^2) \n\\
&= \frac{e^2}{2 \pi^2}\, \frac{\ep^2 - m^2}{\ep^2 (\alpha_0 - \alpha_2) + m^2 (3 \alpha_0 - 4 \alpha_1 + \alpha_2)},
\label{sigmaxx}
\end{align}
that provides the final result for the longitudinal conductivity $\sigma_{xx}=\sigma_{yy}={\cal O}(\alpha_n^{-1})$ in the leading order with respect to the disorder strength, i.\,e. in the so-called Drude approximation.

\subsection{Hall conductivity: Contributions of non-crossing diagrams}
\label{subsecNC}

Replacing full two-band Green's functions with the projected ones $G_+^{R,A}(\bb{p})$ in the expression for $\sigma_{xx}$ provides the correct result for $\sigma_{xx}$ in the order ${\cal O}(\alpha_n^{-1})$, which is the leading order for the longitudinal conductivity. The same procedure [corresponding to diagrams in Fig.~\ref{diagrams}(a) and (e)], gives, however, a vanishing result for $\sigma_{xy}$, since the projected Green's function does not contain information on the time-reversal symmetry breaking in the model. As the result the AHE is sub-leading with respect to $\sigma_{xx}$, which is generally the case for a metal with vanishing single-impurity skew-scattering (i.\,e. for spin-independent disorder). For the AHE virtual processes involving states that are far away from the Fermi surface become important. To get the NCA part of the leading order result for $\sigma_{xy}$ one should simply replace the projected Green's function (\ref{proj}) with the full Green's function (\ref{full}) exactly once in each possible place in each of the ladder diagrams and sum up the results \cite{Sinitsyn2007,Sinitsyn2008rev}. This yields the NCA diagrams illustrated in Fig.~\ref{diagrams}(b), where both dressed vertices are still calculated on the mass shell using projected Green's functions $G_+^{R,A}(\bb{p})$, see Fig.~\ref{diagrams}(e), while thick lines denote the full Green's functions $G^{R,A}(\bb{p})$ given by Eq.~(\ref{full}). Note that in fact it is sufficient to keep a single unprojected Green's function in either upper (retarded) or lower (advanced) part of the diagram in Fig.~\ref{diagrams}(b) and then sum up the results. The diagrams in Fig.~\ref{diagrams}(b) with two full Green's functions is just a convenient way to collect the off-shell contributions from both retarded and advanced sectors.

The sum of NCA diagrams shown in Figure \ref{diagrams}(b) gives rise to the leading-order contribution to the AHE conductivity
\be\label{nc1}
\sigma^\text{nc}_{xy}=\frac{e^2}{2\pi} \int \frac{d^2 p}{(2\pi)^2} \tr  \bar \sigma_x G^R(\bb{p}) \bar \sigma_y G^A(\bb{p}),
\e
which is calculated below. Another contribution to $\sigma_{xy}$, which correspond to the diagrams in Fig.~\ref{diagrams}(c) and (d) not included in the NCA, is, however, of the same order. We postpone its analysis to the next sub-section. 

It is convenient to start the computation of $\sigma^\text{nc}_{xy}$ of Eq.~(\ref{nc1}) with the evaluation of the dressed current vertices $\bar \sigma_{x,y}$ using projected Green's functions. These matrix vertices are given by
\begin{align}
\label{matrixVertex}
\bar \sigma_{x,y}=&\;\sigma_{x,y}+\int d\phi' \alpha(\phi-\phi')J_{x,y}(\phi'),\\
\n &J_x(\phi) = p_0 \tau_\text{tr} | \phi \ra \cos\phi \la \phi |,\\
\n &J_y(\phi) = p_0 \tau_\text{tr} | \phi \ra \sin\phi \la \phi |,
\end{align}
that are readily reconstructed from Eqs.~(\ref{projVertex}-\ref{J}). Explicit calculation of the integrals in Eq.~(\ref{matrixVertex}) gives rise to the following results
\begin{align}
\bar \sigma_x(\phi)=
&\;\sigma_x+(\pi p_0\tau_\text{tr}/2\ep)\lt[2 \alpha_1 (\ep + m \sigma_z)  \cos\phi\rt.\n\\
&\lt.+\,p_0 \alpha_0 \sigma_x+p_0 \alpha_2 (\sigma_x \cos 2\phi  + \sigma_y \sin 2\phi )\rt],\\
\bar \sigma_y(\phi)=
&\sigma_y+(\pi p_0\tau_\text{tr}/2\ep)\lt[2 \alpha_1 (\ep + m \sigma_z)  \sin\phi\rt.\n\\
&\lt.+\,p_0 \alpha_0 \sigma_y+p_0 \alpha_2 (\sigma_x \sin 2\phi  - \sigma_y \cos 2\phi )\rt],
\end{align}
where the transport time is given by Eq.~(\ref{ttr}). Performing the remaining integrations in Eq.~(\ref{nc1}) we obtain the final result for the NCA contribution to the anomalous Hall conductivity,
\begin{align}
\sigma_{xy}^\text{nc}  = &-\frac{e^2}{2\pi}\, 4 \ep m (\alpha_0 - \alpha_1)\n\\
 & \times\frac{ \ep^2 (\alpha_0 - \alpha_2) + m^2 (\alpha_0 - 2 \alpha_1 + \alpha_2)}
      {[\ep^2 (\alpha_0 - \alpha_2) + m^2 (3 \alpha_0 - 4 \alpha_1 + \alpha_2)]^2},
 \label{nc}
\end{align}
that generalizes the NCA contribution obtained in Ref.~\onlinecite{Sinitsyn2006} for the limit of uncorrelated (white noise) disorder, $\alpha_n=\alpha \delta_{n,0}$, which is discussed in Sec.~\ref{subsecWN} in more detail.

The set of diagrams in Fig.~\ref{diagrams}b contains a single diagram with no disorder lines. This diagram represents the intrinsic Fermi surface contribution to the anomalous Hall conductivity \cite{Sinitsyn2006}
\be\label{int}
\sigma^\text{int}_{xy}=\frac{e^2}{2\pi} \int \frac{d^2 \bb{p}}{(2\pi)^2} \tr \sigma_x G^R_0(\bb{p}) \sigma_y G^A_0(\bb{p})=-\frac{e^2 m}{4\pi\ep},
\e
that is manifestly independent of disorder parameters $\alpha_n$.  Note that the value of $\sigma^\text{int}_{xy}$ at $\ep=m$ matches the value of $\sigma_{xy}^\text{II} = -e^2/4\pi$ in the gap $|\ep|<m$ given by the Fermi sea contribution (\ref{II}). The intrinsic conductivity (\ref{int}) can be measured independently at a finite frequency $\tau_\text{tr}^{-1}\ll\omega\ll m,\,\ep$ that is sufficiently large to exceed the relevant disorder scattering rates. In such a high-frequency limit extrinsic contributions, which are sensitive to disorder, become parametrically small.

The extrinsic part of $\sigma_{xy}^\text{nc}$ is given by the difference between the results of Eqs.~(\ref{nc}) and (\ref{int}), $\sigma_{xy}^\text{ext-nc}=\sigma_{xy}^\text{nc}-\sigma^\text{int}_{xy}$ - the auxiliary quantity that cannot, however, be measured in any experiment as the matter of principle. In the next subsection we consider the remaining extrinsic contributions given by X and $\Psi$ diagrams beyond the NCA, which have to be added to $\sigma_{xy}^\text{nc}$ to obtain the complete expression for $\sigma_{xy}$ in the leading zeroth order with respect to the disorder strength.

\subsection{Hall conductivity: Contributions of X and $\bb{\Psi}$ diagrams}
\label{subsecXPsi}

From technical point of view, the X and $\Psi$ diagrams depicted in Fig.~\ref{diagrams}(c) and (d) consist of two dressed vertices (\ref{J}), two crossed impurity lines, and two Green's functions. These diagrams correspond to the following two contributions to the AHE conductivity
\beml
\label{XPSI}
\begin{align}
\sigma_{xy}^\text{X}& = \frac{e^2}{2\pi}\int d\phi_1\, d\phi_2\int \frac{d^2 \bb{p}_3 d^2 \bb{p}_4}{(2\pi)^2}\, \delta(\bb{p}_1 + \bb{p}_2 - \bb{p}_3 - \bb{p}_4)\,\n\\
&\times\alpha_{1,3} \alpha_{2,3}\, \tr \lt( J^x_1 G^R_3 J^y_2 G^A_4 \rt),\label{Xcont}\\
 \sigma_{xy}^\Psi& = \frac{e^2}{2\pi}\int d\phi_1\, d\phi_2\int \frac{d^2 \bb{p}_3 d^2 \bb{p}_4}{(2\pi)^2}\, \delta(\bb{p}_1 - \bb{p}_2 - \bb{p}_3 + \bb{p}_4)\,\n\\
   & \times\alpha_{1,2} \alpha_{1,3}\, \tr \lt( J^x_1 G^R_3 G^R_4 J^y_2 + J^x_1 J^y_2 G^A_4 G^A_3 \rt),\label{Psicont}
\end{align}
\eml
where we use the short-hand notations $\alpha_{i,j} = \alpha(\bb{p}_i - \bb{p}_j)$, $J^x_j = J_x(\phi_j)$, $G^{R,A}_j = G^{R,A}(\bb{p}_j)$. The structure of the integrals is such that momenta $\bb{p}_{1}$ and $\bb{p}_{2}$ are bound to the Fermi surface, while the momenta $\bb{p}_{3}$ and $\bb{p}_{4}$ span the entire momentum space.

From physics point of view the contribution of X and $\Psi$ diagrams take into account an essential part of the full scattering cross-section on a pair of closely located impurities \cite{Ado15}. It will be clear from the analysis of the integrals that the characteristic distance between these impurities is of the order of the Fermi wavelength, hence such a pair represents a rare impurity fluctuation. Nevertheless the contribution to the skew-scattering from such a fluctuation is so large that is has to be taken into account in the leading order with respect to the disorder strength. Only part of this cross-section has been included in the NCA result $\sigma_{xy}^\text{ext-nc}$. One cannot, however, think of an experiment that may differentiate between these two parts of the full scattering cross-section \cite{Ado15,Ado16}. Consequently, only the sum $\sigma_{xy}^\text{ext-nc}+\sigma_{xy}^\text{X}+\sigma_{xy}^\Psi$ corresponds to an experimentally measurable quantity.

In order to take the integrals in Eqs.~(\ref{XPSI}) we first average the integrands with respect to simultaneous rotation of all four momenta. This is equivalent to averaging with respect to the following rotations of current operators, $J_x \mapsto J_x \cos\phi + J_y \sin\phi$ and $J_y \mapsto J_y \cos\phi - J_x \sin\phi$. Such averaging gives rise to the equivalent symmetrized form of the Hall conductivity $\sigma_{xy} \mapsto (\sigma_{xy} - \sigma_{yx})/2$. Relabeling momenta as $\bb{p}_1 \leftrightarrow \bb{p}_2$ and $\bb{p}_3 \leftrightarrow \bb{p}_4$ we cast the integrals in Eqs.~(\ref{XPSI}) in the following form
\beml
\label{XPSI2}
\begin{align}
\sigma_{xy}^\text{X} &= \frac{e^2}{4\pi}\int d\phi_1\, d\phi_2\int\frac{d^2 \bb{p}_3 d^2 \bb{p}_4}{(2\pi)^2}\, \delta(\bb{p}_1 + \bb{p}_2 - \bb{p}_3 - \bb{p}_4)\,\n\\
    &\times\alpha_{1,3} \alpha_{2,3}\, \tr \bigl( J^x_1 G^R_3 J^y_2 G^A_4 - J^x_1 G^A_3 J^y_2 G^R_4 \bigr), \label{Xint}\\
 \sigma_{xy}^\Psi &= \frac{e^2}{4\pi}\int d\phi_1\, d\phi_2\int \frac{d^2 \bb{p}_3 d^2 \bb{p}_4}{(2\pi)^2}\, \delta(\bb{p}_1 - \bb{p}_2 - \bb{p}_3 + \bb{p}_4)\,\n\\
   & \times\alpha_{1,2} \alpha_{1,3}\, \tr \bigl( J^x_1 ( G^R_3 G^R_4 - G^A_3 G^A_4 ) J^y_2\n\\
    &\qquad\qquad\qquad - J^x_1 J^y_2 ( G^R_4 G^R_3 - G^A_4 G^A_3 ) \bigr),
    \label{Psiint}
\end{align}
\eml
where we use the same short-handed notations as in Eqs.~(\ref{XPSI}).

In order to compute the integrals in Eqs.~(\ref{XPSI2}) to the leading (zeroth) order in disorder strength it is legitimate \cite{Ado15} to neglect self-energy in the Green's functions $G_3$ and $G_4$ by replacing them with the corresponding bare Green's functions
\begin{align}
&G^{R,A}_0(\bb{p}) = \frac{N(\bb{p})}{D^{R,A}(\bb{p})},\\
& N(\bb{p}) = \ep + m \sigma_z + \bb{\sigma} \bb{p},\quad
D^{R,A}(\bb{p})= p_0^2 - p^2 \pm i 0,
\end{align}
where there is no distinction between retarded and advanced numerators. At the next step we can use the following identity
\begin{align}
&\frac{1}{D^R_3 D^A_4} - \frac{1}{D^A_3 D^R_4}= \frac{1}{2} \lt( \frac{1}{D^R_3} - \frac{1}{D^A_3} \rt) \lt( \frac{1}{D^R_4} + \frac{1}{D^A_4} \rt)\n\\
&\qquad\qquad\qquad -\frac{1}{2} \lt( \frac{1}{D^R_3} + \frac{1}{D^A_3} \rt) \lt( \frac{1}{D^R_4} - \frac{1}{D^A_4} \rt) \n\\
& = 2\pi i \lt( \frac{\delta(p_3^2 - p_0^2)}{p_4^2 - p_0^2} - \frac{\delta(p_4^2 - p_0^2)}{p_3^2 - p_0^2} \rt),
\label{idRA}
\end{align}
and the closely related identity
\begin{align}
&\frac{1}{D^R_3 D^R_4} - \frac{1}{D^A_3 D^A_4} = \frac{1}{2} \lt( \frac{1}{D^R_3} - \frac{1}{D^A_3} \rt) \lt( \frac{1}{D^R_4} + \frac{1}{D^A_4} \rt)\n\\
&\qquad\qquad\qquad+\frac{1}{2} \lt( \frac{1}{D^R_3} + \frac{1}{D^A_3} \rt) \lt( \frac{1}{D^R_4} - \frac{1}{D^A_4} \rt) \n\\
& = 2\pi i \lt( \frac{\delta(p_3^2 - p_0^2)}{p_4^2 - p_0^2} + \frac{\delta(p_4^2 - p_0^2)}{p_3^2 - p_0^2} \rt),
\label{idRR}
\end{align}
in order to reorganize the integrals in Eq.~(\ref{XPSI2}).

The identities of Eqs.~(\ref{idRA},\ref{idRR}) used in Eq.~(\ref{XPSI2}) bounds one of the momenta $\bb{p}_3$ or $\bb{p}_4$ to the Fermi surface. Applying the transformation $\bb{p}_1 \leftrightarrow \bb{p}_2$ and $\bb{p}_3 \leftrightarrow \bb{p}_4$ to the second term we fix the absolute value $p_3 = p_0$, while the integration over $\bb{p}_4$ is removed due to the momentum-conserving delta function. In this way we obtain
\beml
\label{XPSI3}
\begin{align}
&\sigma_{xy}^\text{X}= \frac{i e^2}{16\pi^2}\int\!\!\!\int\!\!\!\int\! d\phi_1\, d\phi_2\, d\phi_3\, \alpha_{1,3} \alpha_{2,3}\\
&\times\tr \lt\{ \bigl(J^x_1 N_3 J^y_2 - J^y_1 N_3 J^x_2 \bigr)  \frac{N(\bb{p}_1 + \bb{p}_2 - \bb{p}_3)}{(\bb{p}_1 + \bb{p}_2 - \bb{p}_3)^2 - p_0^2}\rt\},\n\\
&\sigma_{xy}^\Psi = \frac{i e^2}{16\pi^2}\int\!\!\!\int\!\!\!\int\! d\phi_1\, d\phi_2\, d\phi_3\, \alpha_{1,2} \alpha_{1,3}\, \label{PSI3}\\
& \times\tr  \lt\{\lt( \bigl( J^y_2 J^x_1 - J^x_2 J^y_1 \bigr) N_3 +  N_3 \bigl( J^y_1 J^x_2 - J^x_1 J^y_2 \bigr) \rt)\rt.\n\\
& \lt.\times\frac{N(\bb{p}_2 + \bb{p}_3 - \bb{p}_1)}{(\bb{p}_2 + \bb{p}_3 - \bb{p}_1)^2 - p_0^2}\rt\},\n
\end{align}
\eml
where all integrals are now bound to the Fermi surface, hence the results are expressed entirely via the harmonics of disorder correlation function defined in Eq.~(\ref{alphaphi}). The traces in Eqs.~(\ref{XPSI3}) are readily computed using Eq.~(\ref{J}) and the identities
\beml
\label{Nid}
\begin{align}
&N(\bb{p}_1 + \bb{p}_2 - \bb{p}_3) = N_1 + N_2 - N_3,\\
&N(\bb{p}_2 + \bb{p}_3 - \bb{p}_1)= N_2 + N_3 - N_1,
\end{align}
\eml
where we defined $N_i= 2\ep | \phi_i \ra \la \phi_i |$.

For the X diagram we further symmetrize the integrand with respect to the replacement $\phi_1 \leftrightarrow \phi_2$ and put $\phi_3 = 0$ since only relative angles matter. This gives
\begin{align}
\sigma_{xy}^\text{X} &= \frac{i e^2}{4\pi}\, \ep^2 \tau_\text{tr}^2 \int\!\!\!\int\! d\phi_1\, d\phi_2\,
\alpha(\phi_1) \alpha(\phi_2)\, \sin(\phi_1 - \phi_2)\, \n\\
&\times\frac{ \la 0 | \phi_1 \ra \la \phi_1 | \phi_2 \ra \la \phi_2 | 0 \ra -\la 0 | \phi_2 \ra \la \phi_2 | \phi_1 \ra \la \phi_1 | 0 \ra}{1 + \cos(\phi_1 - \phi_2) - \cos\phi_1 - \cos\phi_2}
\end{align}
or, more explicitly,
\begin{align}
\sigma_{xy}^\text{X}=&\frac{e^2}{8\pi \ep}\, m (\ep^2 - m^2) \tau_\text{tr}^2 \n\\
&\times\int\!\!\!\int d\phi\, d\phi'\, \alpha(\phi) \alpha(\phi')\, \lt(1 - \cos(\phi - \phi')\rt), \label{XPhi}
\end{align}
where we have used Eq.~(\ref{state}). In terms of angular harmonics the result of Eq.~(\ref{XPhi}) reads
\begin{multline}
  \sigma_{xy}^\text{X}= \frac{e^2}{2\pi}\, \frac{4\ep m (\ep^2 - m^2)(\alpha_0^2 - \alpha_1^2)}
      {\lt(\ep^2 (\alpha_0 - \alpha_2) + m^2 (3 \alpha_0 - 4 \alpha_1 + \alpha_2)\rt)^2},
      \label{XAlpha}
\end{multline}
where only three first harmonics contribute.

The computation of $\Psi$ diagram is slightly more involved. First, we symmetrize Eq.~(\ref{PSI3}) with respect to $\phi_2 \leftrightarrow \phi_3$ and let $\phi_1 = 0$. As the result we obtain the expression
\begin{align}
\sigma_{xy}^\Psi &= \frac{i e^2}{4\pi}\, \ep^2 \tau_\text{tr}^2 \int\!\!\!\int\! d\phi_2\, d\phi_3\,
\alpha(\phi_2) \alpha(\phi_3)\, (\sin\phi_2 - \sin\phi_3) \n\\
&\times\frac{\la 0 | \phi_3 \ra \la \phi_3 | \phi_2 \ra \la \phi_2 | 0 \ra -\la 0 | \phi_2 \ra \la \phi_2 | \phi_3 \ra \la \phi_3 | 0 \ra}{1 + \cos(\phi_2 - \phi_3) - \cos\phi_2 - \cos\phi_3},
\end{align}
that can be rewritten with the help of Eq.~(\ref{state}) as
\begin{align}
\sigma_{xy}^\Psi  =& -\frac{e^2}{8\pi \ep}\, m (\ep^2 - m^2) \tau_\text{tr}^2 \int\!\!\!\int\! d\phi\, d\phi'\, \alpha(\phi) \alpha(\phi')\n \\
&\times\frac{\lt(1 - \cos(\phi - \phi')\rt) \cos(\phi/2 + \phi'/2)}{\cos(\phi/2 - \phi'/2)},
\label{PsiPhi}
\end{align}
where all harmonics of the disorder correlation function play a role in contrast to the other contributions. Indeed, the expression of Eq.~(\ref{PsiPhi}) can
be cast in the following form
\begin{align}
\sigma_{xy}^\Psi = &-\frac{e^2}{2\pi}\, \frac{ 8\ep m (\ep^2 - m^2)}{\lt(\ep^2 (\alpha_0 - \alpha_2) + m^2 (3 \alpha_0 - 4 \alpha_1 + \alpha_2)\rt)^2}\n\\
&\times\lt(\alpha_0 \alpha_1 +2\s_{n=1}^\infty (-1)^n \alpha_n \alpha_{n+1} \rt),
\label{PsiAlpha}
\end{align}
with the help of angular integrations in Eq.~(\ref{PsiPhi}) that are relegated to Appendix \ref{app_derivation}.

It is easy to see from the result of Eq.~(\ref{PsiAlpha}) that the contribution of the $\Psi$ diagram is zero in the case of isotropic disorder $\alpha_n=\alpha \delta_{n,0}$ as has been found also in Ref.~\onlinecite{Ado15}. For the contribution to be finite one needs to have at least two adjacent scattering harmonics which are both non zero.

It is instructive to combine the contributions of X and $\Psi$ diagrams, $\sigma_{xy}^{\text{X} + \Psi}\equiv\sigma_{xy}^\text{X} + \sigma_{xy}^\Psi$, since the result has a particularly simple form
\begin{align}
&\sigma_{xy}^{\text{X} + \Psi}= \frac{e^2}{2\pi}\, \frac{4\ep m (\ep^2 - m^2)}{\lt(\ep^2 (\alpha_0 - \alpha_2) + m^2 (3 \alpha_0 - 4 \alpha_1 + \alpha_2)\rt)^2}\n\\
&\quad\times\lt((\alpha_0 - \alpha_1)^2 +2\s_{n=1}^\infty (-1)^n (\alpha_n - \alpha_{n+1})^2 \rt),
 \label{XPsi}
\end{align}
which is readily evaluated for different disorder models.

\subsection{General result for the AHE}

Before considering specific examples of disorder potentials we shall summarize the final results for the anomalous Hall conductivity $\sigma_{xy}$ in the leading (zeroth) order in disorder strength.

In the insulating gap region $|\ep|<m$, the anomalous Hall conductivity is given by the Fermi sea intrinsic contribution $\sigma_\text{xy}^\textrm{II}=-e^2/4\pi$ of Eq.~(\ref{II}) \cite{Haldane1988}.

In the metallic regime $\ep>m$, the contribution of $\sigma_{xy}^\text{II}={\cal O}(\alpha_n)$ becomes negligible in comparison to the Fermi surface contributions $\sigma_{xy}^\text{I}={\cal O}(\alpha_n^0)$. The total anomalous Hall conductivity above the gap reads
\be\label{sigmaxy}
\sigma_{xy}=\sigma_{xy}^\text{I}=\sigma_{xy}^\text{nc}+\sigma_{xy}^{\text{X} + \Psi},
\e
where the contribution of the non-crossing ladder diagrams in Fig.~\ref{diagrams}(b) is given by Eq.~(\ref{nc}), while the combined contribution of $\text{X}$ and $\Psi$ diagrams in Figs.~\ref{diagrams}(c) and (d) is given by Eq.~(\ref{XPsi}).

Alternatively, $\sigma_{xy}^\text{I}$ can be represented as a sum of intrinsic and extrinsic contributions as
\be\label{intext}
\sigma_{xy}=\sigma^\text{int}_{xy}+\sigma^\text{ext}_{xy},\quad\sigma^\text{ext}_{xy}=\sigma_{xy}^\text{ext-nc}+\sigma_{xy}^{\text{X} + \Psi},
\e
where  $\sigma^\text{int}_{xy}$ is given by Eq.~(\ref{int}) and $\sigma_{xy}^\text{ext-nc}=\sigma_{xy}^\text{nc}-\sigma^\text{int}_{xy}$.

The extrinsic part $\sigma^\text{ext}_{xy}$ can be further divided into the side-jump and skew-scattering contributions. Within the present formalism such a division looks grossly redundant and artificial since it provides us with no additional physical insight.  The side-jump and skew-scattering contributions are indistinguishable parametrically and cannot be measured independently in transport experiments. Therefore, we do not make such a division in the main text of this paper.

On the other hand, the side-jump and skew-scattering contributions appear naturally if one performs the calculations in the eigen basis of the clean Hamiltonian or constructs a generalized Boltzmann equation approach that refers to the eigen basis of the clean Hamiltonian \cite{Sinitsyn2006,Sinitsyn2007,Sinitsyn2008rev}. In Appendix~\ref{app_separate} we provide separate expressions for the side-jump and skew-scattering parts of $\sigma_{xy}^\text{ext-nc}$. Within this classification, $\text{X}$ and $\Psi$ diagrams represent a part of skew-scattering contribution on the pairs of closely positioned impurities, which is not captured within the NCA and was overlooked in studies of the AHE preceding Refs.~\onlinecite{Ado15, Ado16}.

\section{Limiting cases}
\label{secLimits}

\subsection{White noise disorder}
\label{subsecWN}

\noindent
To the best of our knowledge all previous calculations of the anomalous Hall conductivity in metals have been focused on the case of uncorrelated (white-noise) disorder, $\tilde{\alpha}(\bb{r})\propto \delta(\bb{r})$, which corresponds to an isotropic correlator $\alpha(\phi)=\text{const}$. In particular, for massive Dirac fermions the corresponding result for the NCA contribution $\sigma_{xy}^{\text{nc}}$ was obtained in Ref.~\onlinecite{Sinitsyn2006}, while the full anomalous Hall conductivity $\sigma_{xy}$ in the leading order with respect to disorder strength has been computed for the first time in Ref.~\onlinecite{Ado15}.

The white-noise disorder $\alpha(\phi)=\alpha_0$ is characterized by a single Fourier component $\alpha_0$ in Eq.~(\ref{alphaphi}). In this case the NCA contribution of Ref.~\onlinecite{Sinitsyn2006} is readily reproduced from Eq.~(\ref{nc}) as
\be
\label{white_nc}
\sigma_{xy}^\text{nc} = -\frac{e^2}{2\pi}\, \frac{4 \ep m (\ep^2 + m^2)}{(\ep^2 + 3 m^2)^2},
\e
which is manifestly independent of the white-noise disorder strength $\alpha_0$ and decays as $m/\ep$ for $\ep\gg m$. The additional contribution of the crossed X and $\Psi$ diagrams defined by Eq.~(\ref{XPsi}) reads
\begin{equation}
\sigma_{xy}^{\text{X}+\Psi} = \frac{e^2}{2\pi}\, \frac{4\ep m (\ep^2 - m^2)}{(\ep^2 + 3 m^2)^2},
\end{equation}
in agreement with Ref.~\onlinecite{Ado15}. Note that this contribution is similarly independent of $\alpha_0$ and decays as $m/\ep$. One can also see that $\sigma_{xy}^{\text{X}+\Psi}$ and $\sigma_{xy}^\text{nc}$ have opposite signs leading to a reduced value \cite{Ado15} of the total $\sigma_{xy}$ with respect to the NCA result \cite{Sinitsyn2006}. Indeed, the total Hall conductivity
\begin{equation}
\sigma^\text{white noise}_{xy} = -\frac{e^2}{2\pi}\, \frac{8 \ep m^3} {(\ep^2 + 3 m^2)^2},
\label{whitenoise}
\end{equation}
demonstrates a much faster decay $(m/\ep)^3$ in the metal limit $\ep\gg m$. We will see that such an essential cancellation of the AHE at large energies is a special property of the white-noise disorder.

\begin{figure}
\includegraphics[width=\columnwidth]{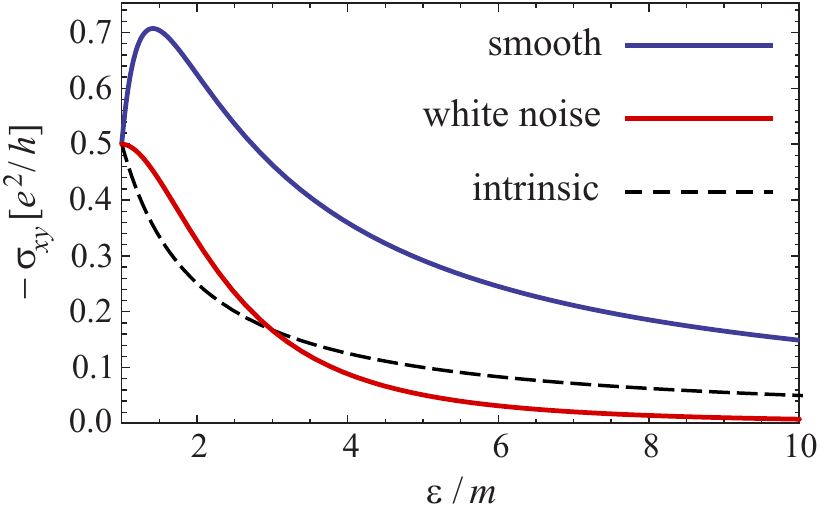}\\
\caption{The anomalous Hall conductivity $\sigma_{yx}=-\sigma_{xy}$ in the units of $e^2/2\pi\hbar$ as a function of Fermi energy, $\ep>m$, above the gap. The lower (red) solid line illustrates the result of Eq.~(\ref{whitenoise}) for the white-noise disorder $\alpha(\phi)=\alpha_0$. The upper (blue) solid line corresponds to the result of Eq.~(\ref{smooth}) for the case of smooth disorder defined in Eq.~(\ref{b}). Dashed line refers to the intrinsic contribution which is given by $\sigma_{xy}^\text{int}$ in Eq.~(\ref{int}) in the absence of disorder.
}
\label{fig_WNSmIn}
\end{figure}

The anomalous Hall conductivity of Eq.~(\ref{whitenoise}) is illustrated in Fig.~\ref{fig_WNSmIn} for the case of short-range white-noise disorder, $\alpha(\phi)=\alpha_0$, by red (lowest solid) line. The intrinsic contribution $\sigma^\text{int}_{xy}=-{e^2 m}/{4\pi\ep}$ from Eq.~(\ref{int}), which would remain in the absence of disorder, is plotted in the same figure with the dashed line. Note that none of the curves actually depend on the disorder parameter $\alpha_0$. For $\ep\gg m$ the extrinsic and intrinsic contributions almost cancel each other such that the corresponding $\sigma_{xy}$ decays as $(m/\ep)^3$ instead of the inverse linear decay of intrinsic $\sigma^\text{int}_{xy}$ and extrinsic $\sigma^\text{ext}_{xy}$ contributions.

The monotonous decay of $\sigma_{xy}$ with energy $\ep$, as well as the cancellation of the leading terms $\propto m/\ep$ between intrinsic and extrinsic contributions at high energies, are not generic features of a weakly disordered system, but rather a specific property of the short range white-noise disorder corresponding to isotropic correlator $\alpha(\phi)=\alpha_0$. Despite $\sigma_{xy}$ is insensitive to the overall strength of a (weak) disorder, i.e., it does not change when all harmonics $\alpha_n$ are simultaneously multiplied by an arbitrary numerical factor, it turns out to be very sensitive to the correlation properties of disorder in general, i.\,e. to the \textit{relative magnitude} of the angular harmonics $\alpha_n$. This interesting property is not at all limited to the model of massive Dirac fermions considered in the present paper, but extends to any metal system where the AHE conductivity is sub-leading (with respect to disorder strength) as compared to the longitudinal one.

%%%%%%%%%%%
%%FIG. 2
%%%%%%%%%%%
%
%%%%%%%%%%%%%%%%%%%%%%%
%%%%%%%%%%%%%%%%%%%%%%%

\subsection{Smooth disorder}
\label{subsecSmooth}
\noindent
The limit of smooth disorder is characterized by small angle scattering with the function $\alpha(\phi)$ peaked around $\phi = 0$. A natural model for such long-range correlated disorder is provided by the angular diffusion. The latter is specified by the angular harmonics in Eq.~(\ref{alphaphi}) of the following form
\beml
\begin{align}
&\alpha_n = a - b n^2,\\
&a  = \int \frac{d\phi}{2\pi}\, \alpha(\phi),\quad
b=\frac{1}{2} \int \frac{d\phi}{2\pi}\, \phi^2 \alpha(\phi) \ll a,
 \label{b}
\end{align}
\eml
where the parameter $a$ represents the forward scattering probability, while the parameter $b$ is the diffusion coefficient in the space of momentum angles (we remind that the absolute value of the momentum is pinned to the Fermi surface).

One observes immediately that the forward-scattering parameter $a$ does enter neither the diagonal nor anomalous Hall conductivity. Indeed, the transport rate of Eq.~(\ref{ttr}), which defines $\sigma_{xx}$ in Eq.~(\ref{sigmaxx}), and both the NCA (\ref{nc}) and the crossed (\ref{XPsi}) contributions to the Hall conductivity in Eq.~(\ref{sigmaxy}) depend only on the angular harmonics differences $\beta_n=\alpha_n-\alpha_{n-1}$ with $n\geq1$. Consequently, the forward scattering parameter $a$ is manifestly canceling out from all transport quantities. We stress, however, that the forward scattering enters explicitly to the quantum scattering rate  $\tau^{-1}=\pi(a\,\ep -b\, p_0^2/\ep)$ as is readily seen from Eq.~(\ref{tau}). This rate describes the decay of a given quantum state which remains finite even in the limit $b\to 0$.

From Eqs.~(\ref{ttr}-\ref{sigmaxx}) we obtain the longitudinal conductivity $\sigma_{xx}\propto \tau_\text{tr}=1/2\pi\ep b$, which is inversely proportional to
the diffusion parameter $b$. In contrast, the Hall conductivity $\sigma_{xy}={\cal O}(\alpha_n^0)$ is again a function of $\ep/m$ only and does not depend explicitly on disorder strength. Upon substitution of Eq.(\ref{b}) into Eq.~(\ref{nc}), we obtain the NCA contribution in the form
\be
\sigma_{xy}^\text{nc} = -\frac{e^2}{2\pi} \lt( \frac{m}{\ep} - \frac{m^3}{2\ep^3} \rt),
\label{nc_smooth}
\e
which is notably different from the corresponding result (\ref{white_nc}) in the case of white-noise disorder.

It is interesting to note that the contributions of X and $\Psi$ diagrams diverge separately as $1/b$ in the limit $b\to 0$, while their combined result is always finite.  Summing up Eqs. (\ref{XPhi}) and (\ref{PsiPhi}) we obtain
\begin{align}
\sigma_{xy}^{\text{X}+\Psi} &= \frac{e^2}{4\pi \ep}\, m (\ep^2 - m^2) \tau_\text{tr}^2 \int\!\!\!\int\! d\phi\, d\phi'\,\alpha(\phi) \alpha(\phi')\n\\
    &\times\frac{\lt(1 - \cos(\phi - \phi')\rt)\sin(\phi/2) \sin(\phi'/2)}{\cos(\phi/2 - \phi'/2)},
 \label{XPsi_phi}
\end{align}
where both angles $\phi$ and $\phi'$ are close to zero due to the peaked character of $\alpha(\phi)$. Consequently it is legitimate to expand the trigonometric functions in the integrand. Using the definition of Eq.~(\ref{b}) we obtain the result
\begin{equation}
 \sigma_{xy}^{\text{X}+\Psi}
  = -\frac{e^2}{2\pi} \lt( \frac{m}{2\ep} - \frac{m^3}{2\ep^3} \rt),
 \label{XPsi_smooth}
\end{equation}
which has to be added to that of Eq.~(\ref{nc_smooth}) to obtain the total Hall conductivity (\ref{sigmaxy}) in the case of smooth disorder (\ref{b}). The result is given by
\be
\sigma^\textrm{smooth}_{xy} = -\frac{e^2}{2\pi}\, \lt( \frac{3 m}{2\ep} - \frac{m^3}{\ep^3} \rt),
\label{smooth}
\e
which behaves very differently from the AHE conductivity in the case of while-noise disorder as illustrated in Fig.~\ref{fig_WNSmIn}. Indeed, unlike in the case of white-noise disorder, the X and $\Psi$ diagrams in the case of long-range disorder enhance the total Hall conductivity. The resulting $\sigma_{xy}$ acquires a non-monotonic dependence on the Fermi energy $\ep$, which is a signature of long-range correlated disorder potential. The result exceeds both the intrinsic contribution (dashed line in Fig.~\ref{fig_WNSmIn}) and the white-noise AHE conductivity [red (lowest solid) line] in the whole range of energies above the gap, $\ep>m$. For $\ep\gg m$, the anomalous Hall conductivity is given by $\sigma_{xy} \approx -(e^2/2\pi)(3m/2\ep)$ such that $\sigma_{xy}=3\sigma_{xy}^\text{int}$ is three times larger than the intrinsic contribution, and $3/2$ times larger than the non-crossing result. Expressions for individual intrinsic, side-jump, and skew-scattering contributions to Eq.~(\ref{smooth}) are given in the Appendix~\ref{app_separate}.

Before concluding the section we show how to obtain the result Eq.~(\ref{XPsi_smooth}) directly from the general formula Eq.~(\ref{XPsi}). Substituting the values of $\alpha_n$ from Eq.~(\ref{b}) we obtain
\be
\sigma_{xy}^{\text{X}+\Psi}= \frac{e^2}{2\pi}\, \frac{m (\ep^2 - m^2)}{4\ep^3}\, \lt( 1 + 2 \sum_{m=1}^\infty (-1)^m (2m + 1)^2 \rt).\n
\e
The sum here is formally divergent but can be nevertheless computed using the identity
\be
\label{zeta}
\sum_{n=1}^\infty \frac{(-1)^n}{(2n + 1)^k} = 2^{-2k} \lt( \zeta(k,5/4) - \zeta(k,3/4) \rt),
\e
where $\zeta(k,p)$ is the Hurwitz zeta function. While the left-hand side of Eq.~(\ref{zeta}) converges only provided $k>0$, its right-hand side can be analytically continued to all complex $k$. At the point $k = -2$ it takes the value $-3/2$ and the result Eq.~(\ref{XPsi_smooth}) is reproduced.

After considering two universal limits of an infinitely short-range and a very long-range disorder, in next section we apply the general results obtained in Sec.~\ref{secGeneral} to two specific disorder models which illustrate possible evolutions of the anomalous Hall conductivity in the region between the above limiting cases.

\section{Crossover from white noise to smooth disorder}
\label{secCrossover}

\subsection{Mixed disorder model}
\label{susecMixed}

The mixed disorder model combines short-range impurities and smooth potential variations. It is characterized by the following harmonics
\be
\alpha_n  = a+u \,\lt((1-w)\delta_{n,0} - w n^2/4\rt) ,\quad 0\leq w\leq 1,
\label{alpha_mixed}
\e
where the parameter $a$ is again responsible for forward scattering, $u$ characterizes the disorder strength, while the parameter $w$ controls the balance between short and long range impurities. From Eqs.~(\ref{ttr}-\ref{sigmaxx}) we obtain the longitudinal conductivity
\be
\sigma_{xx}=\frac{e^2}{2\pi^2 u}\frac{\ep^2-m^2}{\ep^2 + 3 m^2 (1-w)}.
\label{sigmaxx_mixed}
\e
As before, the probability of forward scattering, represented by the parameter $a$, does not affect the transport properties.

Substituting $\alpha_n$ from Eq.~(\ref{alpha_mixed}) into the general expression of Eq.~(\ref{nc}) we obtain
\be
\sigma_{xy}^\text{nc} = -\frac{e^2}{2\pi} \frac{\ep m (4-3w)\lt(\ep^2+m^2(1-3w/2)\rt)}{\lt(\ep^2+3m^2(1-w)\rt)^2},
\label{nc_mixed}
\e
which represents the NCA contribution to the anomalous Hall conductivity for the mixed disorder model. Similarly, from Eq.~(\ref{XPsi}) we obtain
\begin{equation}
\sigma_{xy}^{\text{X}+\Psi} = \frac{e^2}{2\pi} \frac{\ep m (4-6 w +3w^2/2)(\ep^2-m^2)}{\lt(\ep^2+3m^2(1-w)\rt)^2},
\label{XPsi_mixed}
\end{equation}
which represents the contributions of X and $\Psi$ diagrams in the model of Eq.~(\ref{alpha_mixed}).

\begin{figure}
 \includegraphics[width=\columnwidth]{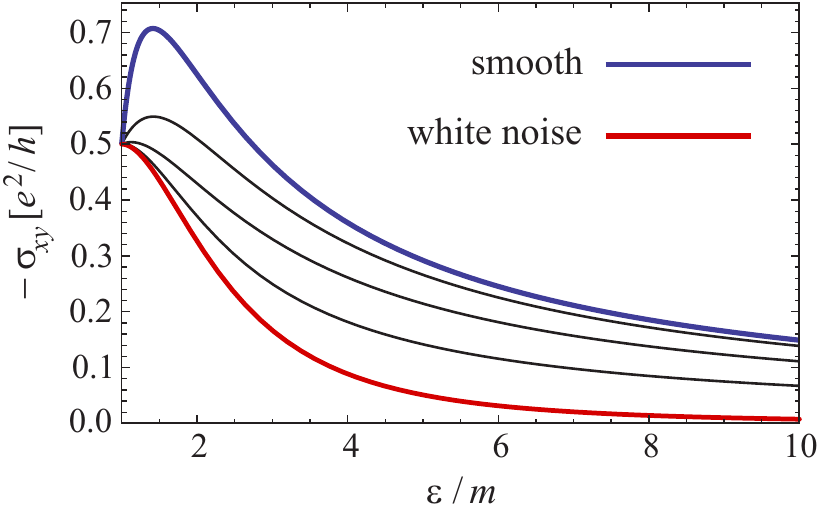}
\caption{
The anomalous Hall conductivity given by Eq.~(\ref{mixed}) for the mixed disorder model (\ref{alpha_mixed}). The curves correspond to different values of $w=\{0,\,0.25,\,0.5,\,0.75,\,1\}$ (from the bottom to the top). The lowest curve ($w=0$) is equivalent to the result of Eq.~(\ref{whitenoise}) for isotropic white-noise disorder scattering, while the highest curve ($w=1$) represents the smooth disorder limit of Eq.~(\ref{smooth}).
}
\label{fig_mixed}
\end{figure}

The sum of Eq.~(\ref{nc_mixed}) and (\ref{XPsi_mixed}) gives rise to the total anomalous Hall conductivity of the form
\begin{equation}
\sigma^\text{mixed}_{xy} = -\frac{e^2}{2\pi}  \frac{\ep m \lt(3\ep^2 w(2-w)+2m^2(8-15w+6w^2)\rt)}{2\lt(\ep^2+3m^2(1-w)\rt)^2},
\label{mixed}
\end{equation}
that is illustrated in Fig.~\ref{fig_mixed} for different values of $w$. The result of Eq.~(\ref{mixed}) is monotonously increasing with $w$ for a given $\ep/m$, so that the smooth disorder limit, $w=1$, corresponds to the largest Hall conductivity as illustrated in Fig.~\ref{fig_mixed}. This property is, however, not universal as we will see in the next Subsection.

In the metallic limit $\ep \gg m$, the Hall conductivity is
\begin{equation}
\sigma^\text{mixed}_{xy}(\ep \gg m) = -\frac{3e^2 m}{4\pi\ep} w(2-w).
\end{equation}
This equation provides a convenient tool to determine the parameter $w$ from experimental data. It is also remarkable that even a small amount of smooth disorder $w > 0$ significantly enhances the AHE in the metallic limit changing $\sigma_{xy}$ from $\propto \ep^{-3}$ to $\propto \ep^{-1}$.

%%%%%%%%%%%
%% FIG. 3
%%%%%%%%%%%
%%
%%%%%%%%%%%%%%%%%%%%%%%
%%%%%%%%%%%%%%%%%%%%%%%

\subsection{Gaussian correlation function}
\label{subsecGaussian}

In this Subsection we propose a particular single-component disorder model that gives rise to a non-monotonous behavior of the anomalous Hall conductivity in the crossover from short-range to smooth disorder. Specifically, we assume a Gaussian correlation function
\be
\alpha(\mathbf{p}) = u \exp\lt(-c^2 p^2/2\rt),
\label{Gauss}
\e
where $c$ is the disorder correlation length. Upon projecting on to the Fermi surface, the angular scattering amplitude and its harmonics are given by
\begin{gather}
\alpha(\phi)=\eta\,  e^{\xi\cos \phi}, \qquad \alpha_n=\eta I_n(\xi),
\label{alpha_gaussian} \\
\eta = u\, e^{-\xi}, \qquad \xi = c^2 (\ep^2 - m^2),
\label{xi}
\end{gather}
where the parameter $\xi$ interpolates between the white-noise ($\xi=0$) and smooth ($\xi \gg 1$) disorder limits as the energy $\ep$ is increased.  In the case $\xi \gg 1$ (which inevitably occurs at large enough $\ep$, since any disorder is smooth on the scale of vanishing Fermi wave length) one may also approximate  $\alpha(\phi)\propto e^{-\xi \phi^2/2}$ for the relevant scattering angles $\phi\lesssim \xi^{-1/2}\ll 1$.

The angular moments of the scattering cross section (\ref{alpha_gaussian}) are given by the modified Bessel functions of the first kind $I_n(\xi)$. The diagonal conductivity $\sigma_{xx}$ in the leading order with respect to disorder strength is  readily found from Eqs.~(\ref{ttr}-\ref{sigmaxx}) as
\be
\sigma_{xx}=\frac{e^2}{2\pi^2\eta}\frac{\ep^2-m^2}{\ep^2(I_0-I_2) + m^2(3 I_0-4 I_1+I_2)},\n
\label{sigmaxx_gaussian}
\e
where the argument $\xi$ of the Bessel functions is omitted for brevity.

\begin{figure}
\includegraphics[width=\columnwidth]{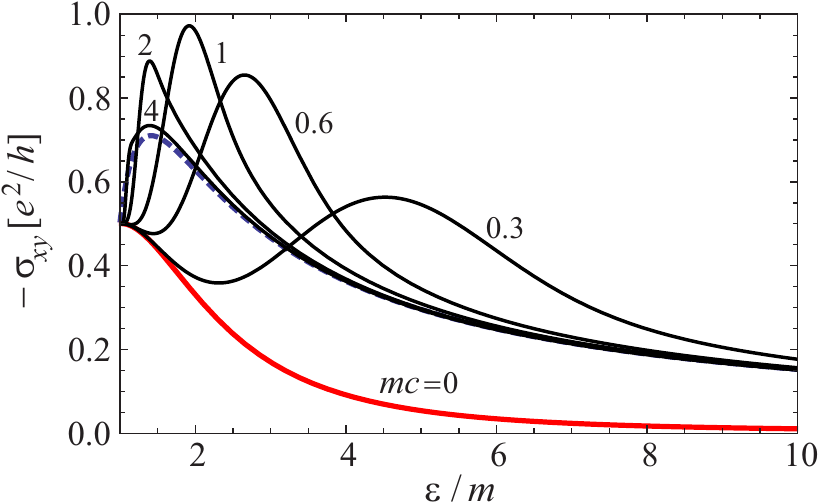}\\
\caption{The anomalous Hall conductivity for the Gaussian disorder model (\ref{Gauss}). The lowest (red) curve shows the result of Eq.~(\ref{whitenoise}) corresponding to the limit of vanishing correlation length $c = 0$ (white-noise disorder). Black solid curves correspond to $m c = \{0.3,\,0.6,\,1,\,2,\,4\}$ as marked. The dashed line corresponds to the smooth disorder limit (\ref{smooth}) of $m c \to \infty$, see also Eq.~(\ref{gauss_asympt}). At intermediate $m c \sim 1$ one clearly observes a transition from the white-noise behavior at low energies $\ep \simeq m$ to the smooth-disorder behavior at large energies $\ep\gg m$, which is manifestly non-monotonic.}
\label{fig_gaussian}
\end{figure}

The infinite series in the expression of Eq.~(\ref{PsiAlpha}) for the $\Psi$ diagram is summed up to the result
\be
\label{Psi_mixed}
\sigma_{xy}^\Psi = -\frac{e^2}{2\pi}\, \frac{8\ep m (\ep^2 - m^2)\lt(\xi (I_0^2 - I_1^2) - I_0 I_1\rt)}
{\lt(\ep^2 (I_0 - I_2) + m^2 (3 I_0 - 4 I_1 + I_2)\rt)^2},
\e
where we have used the properties of the modified Bessel functions. The expression of Eq.~(\ref{Psi_mixed}) can be also obtained directly from the angular integration in Eq.~(\ref{PsiPhi}). The remaining contributions to the anomalous Hall conductivity $\sigma_{xy}=\sigma_{xy}^\text{nc}+\sigma_{xy}^\text{X}+\sigma_{xy}^\Psi$ are given by Eqs.~(\ref{nc}) and (\ref{XAlpha}) with $\alpha_n=\eta I_n(\xi)$. Note that the argument $\xi$ of the modified Bessel functions is energy-dependent, cf.\ Eq.~(\ref{xi}). The final result $\sigma^\text{Gauss}_{xy}$, which does not depend on $\eta$, is illustrated in Fig.~\ref{fig_gaussian} for different values of the correlation length $c$. It is seen that the energy dependence of $\sigma_{xy}$ is in general non-monotonic. In particular, the analysis of the complete expression for $\sigma^\text{Gauss}_{xy}$ shows that the anomalous Hall conductivity reaches its maximum value $\sigma_{yx}\simeq 0.98\,e^2/2\pi$ for $mc \simeq 1.3$ and $\ep/m \simeq 1.8$.

Similar to the case of the mixed disorder model (\ref{alpha_mixed}), the results for the Gaussian model reproduce Eq.~(\ref{whitenoise}) in the white-noise limit $c=0$, see the lowest (red) curve in Fig.~\ref{fig_gaussian}. For any finite correlation length $c > 0$, disorder eventually becomes long-range with increasing energy since the Fermi wave length decreases. Hence the Hall conductivity attains its smooth disorder limit Eq.~(\ref{smooth}) at $\ep \gg \max\{c^{-1}, m\}$. Contrary to the mixed disorder model, the asymptotic value of $\sigma_{xy}$ at large energies is approached either from above or from below depending on the dimensionless parameter $mc$. Indeed, the series expansion of $\sigma_{xy}$ yields
\be
\sigma^\text{Gauss}_{xy} = -\frac{e^2}{2\pi} \lt[\frac{3m}{2\ep} + \lt( \frac{15}{8mc} - 1 \rt) \frac{m^3}{\ep^3} +\dots\rt],
\label{gauss_asympt}
\e
where we have omitted the high-order terms starting from $(m/\ep)^5$ and assumed $\ep c\gg 1$. The result of Eq.~(\ref{gauss_asympt}) indicates that the crossover dependence on $c$ interpolating between short-range and smooth limits is manifestly non-monotonic. 

We conclude that enhanced probability of scattering on small angles for both mixed and Gaussian disorder models (i) makes the dependence of $\sigma_{xy}$ on the energy non-monotonic, and (ii) yields a parametrically larger value of Hall conductivity $\sigma_{xy}\propto m/\ep$ for large energies $\ep\gg m$ as compared to the white noise limit, $\sigma_{xy}\propto (m/\ep)^3$ \cite{Ado15}.

%%%%%%%%%%%%%
% Fig 4.
%
%%%%%%%%%%%%%%%%%%%%%%%
%%%%%%%%%%%%%%%%%%%%%%%

\section{Summary and conclusions}
\label{secSummary}
Despite a long history of the AHE the effect of disorder correlation on the anomalous Hall conductivity did not receive much attention. It has been demonstrated in several previous works that the anomalous Hall conductivity is very sensitive to rare impurity configurations when two impurities appear on a distance of the of order of the Fermi wave length from each other. The skew-scattering on such rare impurity complexes is greatly enhanced resulting in a non-vanishing leading order contribution to the Hall conductivity. This physical picture suggests that, quite generally, the disorder correlations must influence the AHE to a great extent. In this paper we demonstrate by the direct computation that the anomalous Hall conductivity in the model of massive Dirac fermions is indeed highly sensitive to the shape of the disorder correlation function, despite it does not depend on the integral disorder strength. Thus, the AHE can be used to characterize the presence of long-range disorder correlations in some sufficiently clean systems.

The general results for longitudinal and Hall conductivity for arbitrary correlated disorder are presented in Eqs.~(\ref{sigmaxx},\ref{nc},\ref{XPsi},\ref{sigmaxy}). The particular limits of Hall conductivity for the white noise and smooth disorder are given by Eqs.~(\ref{whitenoise},\ref{smooth}). The corresponding results for two interpolating models of mixed and Gaussian-correlated disorder are illustrated in Figs.~\ref{fig_mixed} and \ref{fig_gaussian}.

\begin{acknowledgments}
The work was supported by the Dutch Science Foundation NWO/FOM 13PR3118 and by the Russian Science Foundation under Project 17-12-01359 (M.\,T.). I.\,A.\,D. gratefully acknowledges the support from the Deutsche Forschungsgemeinschaft (project No. DM 1/4-1).
\end{acknowledgments}

\appendix

\section{Derivation of Eq.~(\ref{PsiAlpha})}
\label{app_derivation}

To evaluate the angular integral in Eq.~(\ref{PsiPhi}) we use the Fourier representation for the correlator $\alpha(\phi)$
\begin{align}
F&=\!\!\int\!\!\!\int\!\! d\phi d\phi' \alpha(\phi) \alpha(\phi') \frac{\lt(1 - \cos(\phi - \phi')\rt) \cos(\phi/2 + \phi'/2)}{\cos(\phi/2 - \phi'/2)}\n\\
 &=\alpha_0^2F_{00}+2\alpha_0\s_{k=1}^\infty \alpha_k(F_{0k}+F_{k0})+4\s_{l,k=1}^\infty\alpha_l\alpha_kF_{lk},
\label{Fourier}
\end{align}
where we introduced the harmonics
\begin{align}
F_{lk}=&\int\!\!\!\int\! d\phi\, d\phi'\, \cos(l \phi) \cos(k \phi') \n\\
&\times \frac{\lt(1 - \cos(\phi - \phi')\rt)\cos(\phi/2 + \phi'/2)}{\cos(\phi/2 - \phi'/2)} ,
\end{align}
which can be explicitly evaluated. With the help of new variables $\phi_{\pm} =(\phi\pm\phi')/2$ we write
\begin{align}
&F_{lk}= 2\int_{-\pi/2}^{\pi/2}\!\! d\phi_-\, \frac{1 - \cos 2\phi_-}{\cos\phi_-} \\
&\times\int_0^{2\pi}\!\!\! d\phi_+\, \cos[l (\phi_+ + \phi_-)] \cos[k (\phi_+ - \phi_-)] \cos\phi_+,\n
\end{align}
where the integration over $\phi_+$ can be straightforwardly performed. The result is given by
\begin{multline}
\int_0^{2\pi} \!\!\!\! d\phi_+\, \cos(l (\phi_+ + \phi_-)) \cos(k (\phi_+ - \phi_-)) \cos\phi_+ \n\\
= \frac{\pi}{2} \Big\{ \lt( \delta_{l+1,k}+ \delta_{l-1,k} \rt) \cos((l + k)\phi_-) \n\\
+ \lt( \delta_{l+1,-k} + \delta_{l-1,-k} \rt)\cos((l - k)\phi_-)\Big\},
\end{multline}
where $\delta_{l,k}$ is the Kronecker delta. We observe that all factors in the cosine arguments are odd multiples of $\phi_-$. This allows us to use the following identity (for $n \geq 0$)
\be
\frac{\cos[(2 n + 1)\phi_-]}{\cos\phi_- }= (-1)^n \lt\{1 + 2\s_{k=1}^n (-1)^k \cos 2k\phi_-\rt\},\n
\e
in order to remove $\cos\phi_-$ from the denominator. Once the denominator is removed, the integration over $\phi_-$ yields the expression
\be
F_{lk}= 2 \pi^2  \lt( (-1)^l\, \delta_{|l|+1, |k|} + (-1)^k\, \delta_{|l|-1, |k|}\rt),
\e
which is to be substituted into the Fourier expansion of Eq.~(\ref{Fourier}). As the result we obtain
\be
F = 8\pi^2 \lt( \alpha_0 \alpha_1 - 2 \alpha_1 \alpha_2 + 2 \alpha_2 \alpha_3 - 2 \alpha_3 \alpha_4 + \ldots \rt),
\e
that reproduces the expression of Eq.~(\ref{PsiAlpha}).

\section{Separation of Eq.~(\ref{nc}) into intrinsic, side-jump, an skew-scattering contributions}
\label{app_separate}

As mentioned in Sec.~\ref{subsecNC} the ladder diagrams in Fig.~\ref{diagrams}(b) involve both intrinsic and extrinsic contributions to $\sigma_{xy}$. The latter can be further divided into side-jump and skew-scattering contributions, which appear naturally if one performs calculations in the eigen basis of the clean Hamiltonian or within a generalized Boltzmann equation approach \cite{Sinitsyn2006, Sinitsyn2007}. For completeness here we provide separate expressions for the side-jump and skew-scattering parts of $\sigma_{xy}^\text{ext-nc}$ for the case of correlated disorder. For a more detailed discussion of their physical meaning, we refer the reader to Refs.~\cite{Sinitsyn2006,Sinitsyn2007,Sinitsyn2008rev}.

\textit{Intrinsic contribution.} The set of diagrams in Fig.~\ref{diagrams}(b) contains a single diagram which does not involve any impurity scattering (neither the impurity lines dressing the current vertex nor the disorder-induced self-energy in the involved Green's functions). In the eigen basis of the clean Hamiltonian, the current (velocity) operator contains non-diagonal matrix elements. The intrinsic contribution in this basis results from a bare conductivity bubble containing two non-diagonal current vertices connected by one on-shell and one off-shell clean Green's functions \cite{Sinitsyn2006}. These diagrams produce $\sigma_{xy}^\text{int}=-e^2 m/4\pi\ep$, see Eq.~(\ref{int}).

\textit{Side-jump contribution.} The diagrams in Fig.~\ref{diagrams}(b), which involve exactly one (either left or right) non-diagonal current vertex in the eigen basis of the clean Hamiltonian, contribute to the side-jump part of $\sigma_{xy}$
\be\label{sj}
\sigma_{xy}^\text{sj}  = -\frac{e^2}{2\pi}\, \frac{2 m (\ep^2 - m^2) (\alpha_0 - \alpha_1)}{\ep [\ep^2 (\alpha_0 - \alpha_2) + m^2 (3 \alpha_0 - 4 \alpha_1 + \alpha_2)]},
 \e
that is again defined by the three first harmonics of the correlation function. The single off-shell Green's function in such diagrams connects the non-diagonal vertex to the closest impurity line which can be either a part of the self-energy or of the vertex correction. Both the intrinsic and the side-jump parts of $\sigma_{xy}$ are fully captured within the NCA.

\textit{Skew-scattering contribution.} The NCA part of the skew-scattering contributions to the anomalous Hall conductivity
\be\label{skew-nc}
\sigma_{xy}^\text{skew-nc} = -\frac{e^2}{2\pi}\, \frac{m (\ep^2 - m^2)^2 (\alpha_0 - \alpha_2) (3 \alpha_0 - 4 \alpha_1 + \alpha_2)}  {2 \ep [\ep^2 (\alpha_0 - \alpha_2) + m^2 (3 \alpha_0 - 4 \alpha_1 + \alpha_2)]^2},
\e
is obtained from those diagrams in Fig.~\ref{diagrams}(b) that do not involve the non-diagonal vertices. In such diagrams, the off-shell Green's function connects two impurity lines (both of them can be parts of either the self-energy or the vertex correction). Since the off-shell Green's function decays fast in the real space, the characteristic distance between the corresponding scattering events is rather small (of the order of the Fermi wave length). Therefore, such two impurities should be treated as a single quantum object. The corresponding skew-scattering cross section involves quantum interference terms beyond the NCA. These interference terms are represented by X and $\Psi$ diagrams in Fig.~\ref{diagrams}(c) and (d) which thus provide an inherent part of the skew-scattering cross section on pairs of close impurities.

Altogether the contributions of Eqs.~(\ref{int}), (\ref{sj}), and (\ref{skew-nc}) combine into the total NCA contribution $\sigma_{xy}^\text{nc}$, which is given by Eq.~(\ref{nc}). On the other hand the sum $\sigma_{xy}^\text{skew}=\sigma_{xy}^\text{skew-nc}+\sigma_{xy}^{\text{X}+\Psi}$ gives the total skew-scattering contribution including the quantum interference terms %ID originating in the pair of closely located impurities 
$\sigma_{xy}^{\text{X}+\Psi}$ that are omitted in the NCA.

Using the above expressions one readily obtains the side-jump and skew-scattering parts of $\sigma_{xy}$ for arbitrary $\alpha(\phi)$. In the limit of short-range disorder, $\alpha(\phi)=\alpha_0$, the equations above reduce to the results given in Refs.~\onlinecite{Sinitsyn2006} and \onlinecite{Ado15}. In the opposite limit of ultimately long-range disorder, considered in Sec.~\ref{subsecSmooth}, one arrives to the expressions
\begin{align}
&\sigma_{xy}^\text{int} = -\frac{e^2}{2\pi}\, \frac{m}{2 \ep},\qquad
&\sigma_{xy}^\text{sj} = -\frac{e^2}{2\pi} \lt( \frac{m}{2\ep} - \frac{m^3}{2\ep^3} \rt),\n \\
&\sigma_{xy}^\text{skew-nc}= 0,
&\sigma_{xy}^{\text{X}+\Psi} = -\frac{e^2}{2\pi} \lt( \frac{m}{2\ep} - \frac{m^3}{2\ep^3} \rt),\n
%\label{smoothParts}
\end{align}
which indicate that the NCA part of the skew scattering term vanishes in the limit of smooth disorder. The skew scattering is finite only due to the crossed diagrams and equals to the side jump term. At large energies, $\ep\gg m$, the anomalous Hall conductivity scales as $\sigma_{xy} \approx -(e^2/2\pi)(3m/2\ep)$. In this limit, $\sigma_{xy}=3\sigma_{xy}^\text{int}$ is given by the sum of intrinsic, side-jump, and skew scattering contributions that are all equal to each other.

\end{document}